\documentstyle[11pt]{article}
\begin{document}
\title{\Large\bf  A possible assignment for the ground scalar meson nonet}

\author{\small De-Min Li $^{1,2}$\footnote{E-mail: lidm@zzu.edu.cn; lidm@mail.ihep.ac.cn},~
Ke-Wei Wei $^{1}$, ~Hong Yu $^{2}$\\
\small  $^1$ Department of Physics, Zhengzhou University,
Zhengzhou, Henan 450052, P. R. China\footnote{Mailing address}\\
\small  $^2$ Institute of High Energy Physics, Chinese Academy of
Sciences, Beijing 100039, P. R. China\\}
\date{\today}
\maketitle
\vspace{0.5cm}

\begin{abstract}
Based on the main assumption that the $a_0(980)$ and $D^\ast_{sJ}(2317)$ belong to the $1~^3P_0$
$q\bar{q}$ multiplet, in the framework of Regge phenomenology and meson-meson mixing, it is
suggested that the $a_0(980)$, $K^\ast_0(1052)$, $f_0(1099)$ and $f_0(530)$ constitute the ground
scalar meson nonet, and that the $f_0(1099)$ is composed mostly of $s\bar{s}$ while the $f_0(530)$
is mainly $u\bar{u}+d\bar{d}$. It is supposed that these states would likely correspond to the
observed scalar states $a_0(980)$, $\kappa(900)$,
 $f_0(980)$ and $f_0(600)/\sigma$, respectively. The agreement between the present findings and
those given by other different approaches is satisfactory.

\end{abstract}

\vspace{0.5cm}



\newpage

\baselineskip 24pt

\section*{I. Introduction}
\indent \vspace*{-1cm}

 The spectrum and structure of the scalar mesons are
one of the most controversial subjects in hadron physics. In the
recent issue of Review of Particle Physics\cite{pdg}, too many
light scalar mesons in the region below 2 GeV are claimed to exist
experimentally: two isovectors $a_0(980)$ and $a_0(1450)$, five
isoscalars $f_0(600)/\sigma$, $f_0(980)$, $f_0(1370)$, $f_0(1500)$
and $f_0(1710)$; and three isodoublets $K^\ast_0(1430)$,
$K^\ast_0(1950)$ and $K^\ast_0(800)/\kappa$. Among these states,
it is not yet clear which are the members of the ground scalar
meson nonet.

With respect to the nature of the $a_0(980)$, although some possible interpretations such as
$K\bar{K}$ molecule\cite{KK}, four-quark state\cite{4quark} were proposed in the literature, many
results given by different approaches support the argument that the $a_0(980)$ belongs to the
ground scalar meson multiplet: (1) The K-matrix analysis of the $K\pi$ S-wave\cite{kpi} showed the
mass of the $1~^3P_0$ isovector state is about $960\pm 30$ MeV and supported that the $a_0(980)$ is
dominantly $q\bar{q}$ system; (2) The naive quark model predicts that the LS force makes lighter
the $J=0$ states with respect to the $J=2$, which favors that the $a_0(980)$ rather than the
$a_0(1450)$ belongs to the the scalar member of the lowest $^3P_J$ multiplet, because the
$a_2(1320)$ is well established as a $q\bar{q}$ pair. The same behavior is evident in the
$c\bar{c}$ and the $b\bar{b}$ spectra\cite{vij}; (3) Based on the theory of fine structure, it is
suggested that it is the $a_0(980)$ but not the $a_0(1450)$ that could be a candidate for the
ground $^3P_0$ state \cite{bad}; (4) Most of the fits of the data using the nonrelativistic quark
model strongly favored that the $a_0(980)$ is the isovector member of the ground scalar
nonet\cite{nrqm}; (5) The calculation of the partial width for the decay
$a_0(980)(f_0(980))\rightarrow\gamma\gamma$\cite{tworr}
 based on the assumption that the $a_0(980)$ and $f_0(980)$ are the members of
the $1~^3P_0$ $q\bar{q}$ multiplet is in reasonable agreement with
 the experimental data, which supports the idea of $q\bar{q}$ origin of the scalar mesons $a_0(980)$ and $f_0(980)$; (6) The systematics of scalar
 $q\bar{q}$ states
 on the linear trajectories in the ($n, M^2$) and ($J, M^2$) plane
 indicate the $a_0(980)$ lays comfortable on the linear
 trajectory, together with other scalar states\cite{lintraj}; (7) The
 calculation within QCD sum rules method based on the argument
 that $a_0(980)$ is considered as a $q\bar{q}$ bound state is
 consistent with the existing experimental data\cite{sr}; (8) Some theoretical
 models such as U(3)$\times$U(3) $\sigma$
model\cite{nap}, SU(3) $\sigma$ model
 \cite{su3}, chiral quark model of Nambu-Jona-Lasinio type\cite{9904226} also
suggested the mass of the isovector member of the ground scalar nonet is close to that of the
$a_0(980)$.

Recently, the experimental discovery of the low-lying charm-strange meson
$D^\ast_{sJ}(2317)$\cite{dsj} maybe open a new window to reveal the nature of the scalar states.
All experimental findings, such as all the observed decay modes and angular distributions are
consistent with the interpretation as P-wave states with spin-parity assignment $J^P=0^+$ for the
$D^\ast_{sJ}(2317)$. On the one hand, the picture of the $D^\ast_{sJ}(2317)$ composed of a heavy
quark $c$ and a light quark $s$ fits well with the heavy-quark, chiral symmetries that predict
parity doubling states ($0^-$, $1^-$) and ($0^+$, $1^+$), with the inerparity mass splittings in
the chiral limit given by the Goldberger-Treiman relation, the subsequent observation of $1^{+}$
state $D^\ast_{sJ}(2460)$ strongly supports this picture\cite{loop}, and the assignment that the
$D^\ast_{sJ}(2317)$ is the $c\bar{s}$
 member of the $1~^3P_0$ $q\bar{q}$ multiplet has been suggested by Particle Data Group\cite{pdg}. On the
other hand, the $c\bar{s}$ picture of this state does not play well with the potential model
calculations, which generally predict substantially larger mass. For example, the measured mass of
the $D^\ast_{sJ}(2317)$ is $2317.4\pm 0.9$ MeV, while the prediction of the $1~^3P_0$ $c\bar{s}$
state by Isgur and Godrey is 2.48 GeV\cite{p1} and that by Di Pierro and Eichten is 2.487
GeV\cite{p2}, which are about 160 MeV higher than the measured mass of the $D^\ast_{sJ}(2317)$. The
substantially small observed mass led to many other interpretations on the nature of the
$D^\ast_{sJ}(2317)$, such as the $(DK)$ molecule,
 four-quark state, $D\pi$ atom or baryonium.
 For the detailed review see {\sl e.g.}
Refs.\cite{rev1,rev2}. However, it should be noted that
 the one loop chiral corrections for heavy-light mesons in
potential model\cite{loop} and the coupled channel effect\cite{couple} can naturally account for
the unusually mass of the $D^\ast_{sJ}(2317)$, which confirms the $q\bar{q}$ picture of the
$D^\ast_{sJ}(2317)$. More recently, radiative decays of the $D^\ast_{sJ}(2317)$ and
$D^\ast_{sJ}(2460)$ have been studied by Colangelo et al. within light-cone QCD sum rules, the
results show that invoking nonstandard interpretations of the $D^\ast_{sJ}(2317)$ and
$D^\ast_{sJ}(2460)$ is not necessary, and strongly favor the idea of ordinary $c\bar{s}$ origin of
of the $D^\ast_{sJ}(2317)$ and $D^\ast_{sJ}(2460)$\cite{0505195}.

In the present work, we shall assume that the $a_0(980)$ and $D^\ast_{sJ}(2317)$ are the members of
the $1~^3P_0$ $q\bar{q}$ multiplet, and discuss a possible assignment for the ground scalar
$q\bar{q}$ nonet in the framework of Regge phenomenology and meson-meson mixing.

\section*{II. The mass of the $1~^3P_0$ $n\bar{s}$ state in Regge phenomenology}
\indent\vspace*{-1cm}

A series of recent papers\cite{lintraj,lintraj1,dmli} indicate that the quasi-linear Regge
trajectory can, at least at present,
 give a
reasonable description for the meson spectroscopy, and its
predictions may be useful for the discovery of the meson states
which have not yet been observed. By assuming the existence of the
quasi-linear Regge trajectories for a meson multiplet, one can
have
\begin{equation}
J=\alpha_{i\bar{i^\prime}}(0)+\alpha^\prime_{i\bar{i^\prime}} M^2_{i\bar{i^\prime}},
\label{trajectory}
\end{equation}
where $i$ ($\bar{i^\prime}$) refers to the quark (antiquark) flavor, $J$ and $M_{i\bar{i^\prime}}$
are respectively the spin and mass of the $i\bar{i^\prime}$ meson, $\alpha_{i\bar{i^\prime}}(0)$
and $\alpha^\prime_{i\bar{i^\prime}}$ are respectively the intercept and slope of the trajectory on
which the  $i\bar{i^\prime}$ meson lies. For a meson multiplet, the parameters for different
flavors can be related by the following relations(see Ref.\cite{dmli} and references therein)

(i) additivity of intercepts,
\begin{equation}
\alpha_{i\bar{i}}(0)+\alpha_{j\bar{j}}(0)=2\alpha_{j\bar{i}}(0),
\label{intercept}
\end{equation}

(ii) additivity of inverse slopes,
\begin{equation}
\frac{1}{\alpha^\prime_{i\bar{i}}}+\frac{1}{\alpha^\prime_{j\bar{j}}}=\frac{2}{\alpha^\prime_{j\bar{i}}}.
\label{slope}
\end{equation}

From relations (\ref{trajectory})-(\ref{slope}), one can have
\begin{eqnarray}
M^2_{n\bar{s}}=\frac{\alpha^\prime_{n\bar{n}}M^2_{n\bar{n}}-\alpha^\prime_{c\bar{c}}M^2_{c\bar{c}}+2\alpha^\prime_{c\bar{s}}M^2_{c\bar{s}}}{2\alpha^\prime_{n\bar{s}}},
\label{nsmass}
\end{eqnarray}
where $n$ denotes $u$- or $d$-quark.

 In our estimate of the mass of the $1~^3P_0$ $n\bar{s}$ state, we adopt the assumption presented by Ref.\cite{dmli} that the
slopes of the parity partners' trajectories coincide. Under this assumption, the slopes of the scalar meson trajectories are the same
as those of the vector meson trajectories.
 With the help of slopes of the vector meson trajectories extracted by Ref.\cite{dmli}, we have
$\alpha^\prime_{n\bar{n}}=0.8830$ GeV$^{-2}$, $\alpha^\prime_{n\bar{s}}=0.8493$ GeV$^{-2}$,
$\alpha^\prime_{c\bar{c}}=0.4364$ GeV$^{-2}$ and $\alpha^\prime_{c\bar{s}}=0.5692$ GeV$^{-2}$.
Inserting $M_{n\bar{n}}=M_{a_0(980)}=984.7\pm 1.2 $ MeV, $M_{c\bar{c}}=M_{\chi_{c0}(1P)}=3415.19\pm
0.34$ MeV and $M_{c\bar{s}}=M_{D^\ast_{sJ}(2317)}=2317.4\pm 0.9$ MeV\cite{pdg} into relation
 (\ref{nsmass}), one can have $M_{n\bar{s}}=1051.99\pm 1.48$ MeV.

\section*{III. The $1~^3P_0$ meson nonet in meson-meson mixing}
\indent\vspace*{-1cm}

It is well known that in a meson nonet, the pure isoscalar $n\bar{n}$ and $s\bar{s}$ states can mix to produce the physical isoscalar
states $f_0(M_1)$ and $f_0(M_2)$. In order to understand the physical scalar states, we shall discuss the mixing of the $n\bar{n}$
and $s\bar{s}$ states below.

In the $N=(u\bar{u}+d\bar{d})/\sqrt{2}$, $S=s\bar{s}$ basis, the mass-squared matrix describing the
mixing of the $f_0(M_1)$ and $f_0(M_2)$ can be written as\cite{su3,jpg}
\begin{equation}
M^2=\left(\begin{array}{cc}
M^2_N+2\beta&\sqrt{2}\beta X\\
\sqrt{2}\beta X&2M^2_{n\bar{s}}-M^2_N+\beta X^2
\end{array}\right),
\label{mix}
\end{equation}
where $M_N$ and $M_{n\bar{s}}$ are the masses of the states $N$ and $n\bar{s}$, respectively;
 $\beta$ denotes the total
annihilation strength of the $q\bar{q}$ pair for the light flavors $u$ and $d$; $X$ describes the
SU(3)-breaking ratio of the nonstrange and strange quark propagators via the constituent quark mass
ratio $m_u/m_s$. The masses of the two physical scalar states $f_0(M_1)$ and $f_0(M_2)$, $M_1$ and
$M_2$, can be related to the matrix $M^2$ by the unitary matrix $U$
\begin{eqnarray}
UM^2U^\dagger=\left(\begin{array}{cc}
M^2_1&0\\
0&M^2_2
\end{array}\right),
\label{dig}
\end{eqnarray}
and the physical states $f_0(M_1)$ and $f_0(M_2)$ can be expressed as
\begin{eqnarray}
\left(\begin{array}{l}
f_0(M_1)\\
f_0(M_2)
\end{array}\right)=
U\left(\begin{array}{l}
N\\
S
\end{array}\right).
\label{NS}
\end{eqnarray}

The constituent quark mass ratio can be determined within the nonrelativistic constituent quark
model(NRCQM). In NRCQM\cite{nrqm,prd57}, the mass of a $q\bar{q}$ state with $L=0$, $M_{q\bar{q}}$
is given by
\begin{eqnarray}
M_{q\bar{q}}=m_q+m_{\bar{q}}+\Lambda\frac{\mbox{\bf s}_q\cdot\mbox{\bf s}_{\bar{q}}}{m_q
m_{\bar{q}}},\nonumber
\end{eqnarray}
where $m$ and $\mbox{\bf s}$ are the constituent quark mass and spin, $\Lambda$ is a constant.
Since $\mbox{\bf s}_q\cdot\mbox{\bf s}_{\bar{q}}=-3/4$ for spin-0 mesons and $1/4$ for spin-1
mesons, in the SU(2) flavor symmetry limit, one can have\footnote{Here we take $M_{\pi}=134.9766\pm
0.0006$ MeV, $M_\rho=775.8\pm 0.5$ MeV, $M_K=497.648\pm 0.022$ MeV and $M_{K^\ast}=896.10\pm 0.27$
MeV\cite{pdg}.}
\begin{eqnarray}
X\equiv\frac{m_u}{m_s}=\frac{M_{\pi}+3M_{\rho}}{2M_K+6M_{K^\ast}-M_{\pi}-3M_{\rho}}=0.6298\pm
0.0068.\nonumber
\end{eqnarray}

From relation (\ref{dig}), one can have
\begin{equation}
\begin{array}{l}
2M^2_{n\bar{s}}+(2+X^2)\beta=M^2_1+M^2_2,\\
(M^2_N+2\beta)(2M^2_{n\bar{s}}-M^2_N+\beta X^2)-2\beta^2 X^2=M^2_1M^2_2.
\end{array}
\label{trace}
\end{equation}

For the scalar meson nonet, the masses of two isoscalar physical states satisfy the following approximate sum rule
\begin{eqnarray}
M^2_1+M^2_2\simeq2(M^2_K+M^2_{n\bar{s}})-(M^2_{\eta}+M^2_{\eta^\prime}), \label{sumrule}
\end{eqnarray}
which is derived by Dmitrasinovic in the framework of the Nambu-Jona-Lasinio model with a  $U_A(1)$ symmetry-breaking
instanton-induced 't Hooft interaction\cite{sumrule}.

With the help of $M_N=M_{a_0(980)}$ and $M_{n\bar{s}}=1051.99\pm 1.48$ MeV estimated in section II,
from relations (\ref{mix})-(\ref{sumrule}), we can obtain\footnote{We take $M_\eta=547.75\pm 0.12$
MeV, $M_{\eta^\prime}=957.78\pm 0.14$ MeV\cite{pdg}.}
\begin{eqnarray}
M_1\simeq 1099.86\pm 2.71~\mbox{MeV}, ~M_2\simeq 530.67\pm 1.92~\mbox{MeV},~\beta=-(301281.0\pm
165.7)~\mbox{MeV}^2,
\end{eqnarray}
and
\begin{eqnarray}
\left(\begin{array}{c}
f_0(M_1)\\
f_0(M_2)
\end{array}\right)\simeq\left(\begin{array}{ll}
0.303\pm 0.002&-(0.953\pm 0.001)\\
0.953\pm 0.001&~~0.303\pm 0.002
\end{array}\right)\left(\begin{array}{ll}
N\\
S
\end{array}\right).
\label{content}
\end{eqnarray}

Therefore, under the assumption that the $a_0(980)$ and $D^\ast_{sJ}(2317)$ belong to the $1~^3P_0$
meson multiplet, in the Regge phenomenology and meson-meson mixing, we suggest that the $a_0(980)$,
$K^\ast_0(1052)$, $f_0(1099)$ and $f_0(530)$ constitute the ground scalar meson nonet.

\section*{IV. Discussions}
\indent\vspace*{-1cm}

 Obviously, the mass of the $f_0(530)$ agrees with that of the observed scalar
resonance $f_0(600)/\sigma$ with a mass range of 400-1200 MeV, also, the picture that the
$f_0(530)$ is composed mostly of nonstrange quarkonia is consistent with the decay patterns of the
$f_0(600)/\sigma$\cite{pdg}. This suggests that the $f_0(530)$ would correspond to the observed
state $f_0(600)/\sigma$.

The K-matrix analysis of the $K\pi$ S-wave by Anisovich et al.\cite{kpi} reveals the lowest scalar
kaon with the pole position at $1090\pm 40$ MeV, which favors our estimated mass of the
$K^\ast_0(1052)$. Comparison of the $K^\ast_0(1052)$ and the observed scalar kaon states, $\kappa$,
$K^\ast_0(1430)$ and $K^\ast_0(1950)$, indicates that if the $\kappa$ really exists, the
$K^\ast_0(1052)$ would very likely correspond to the $\kappa(900)$ with a mass of $905^{+65}_{-30}$
MeV\cite{ishida}.

With respect to the $f_0(1099)$, its estimated mass is close to the mass of the observed scalar
state $f_0(980)$ ($980\pm 10$ MeV), also close to the mass of the observed scalar state $f_0(1370)$
($1200-1500$ MeV), and relation (\ref{content}) clearly shows that the $f_0(1099)$ is composed
mostly of $s\bar{s}$.  The results of analysis\cite{ani} for the two-meson spectra support the
picture that the $f_0(980)$ is composed mostly of $s\bar{s}$ quarks. The transition
$\phi(1020)\rightarrow \gamma f_0(980)$ can be well described within the approach of additive quark
model, with the dominant $q\bar{q}$ component in the $f_0(980)$\cite{0403123}, and the decay
$f_0(980)\rightarrow\gamma\gamma$ can be also treated in terms of the $q\bar{q}$ structure of the
$f_0(980)$\cite{tworr,tworr1}. The values of partial widths in both decays (
$\phi(1020)\rightarrow\gamma f_0(980)$ and $f_0(980)\rightarrow \gamma\gamma$) support the
existence of a significant $s\bar{s}$-component in  the $f_0(980)$.  The study of the
$D^+_s\rightarrow \pi^+ f_0(980)$ decay by many authors\cite{d1,d2,d3,d4} also led to the
conclusion about the $s\bar{s}$ nature of the $f_0(980)$. The decay patterns of the
$f_0(1370)$\cite{pdg} implies that the $f_0(1370)$ should be mainly non-strange. Therefore, the
mass and the quarkonia content of the $f_0(1099)$ strongly suggest that the $f_0(1099)$ would
correspond to the observed scalar state $f_0(980)$ rather than the $f_0(1370)$.

 Based on the above analysis, the results of the present work
 predict the ground scalar meson nonet consisting of the $a_0(980)$, $K^\ast_0(1052)$, $f_0(1099)$ and
 $f_0(530)$. These states would correspond to the observed scalar states $a_0(980)$, $\kappa(900)$,
 $f_0(980)$ and $f_0(600)/\sigma$, respectively.

The masses of the ground scalar meson nonet has been estimated by Volkov\cite{9904226} in the
framework of a nonlocal version of a chiral quark model of the Nambu-Jona-Lasinio type where the
correct masses for the ground pseudoscalar meson nonet and vector meson nonet can be produced. The
calculation of Volkov\cite{9904226} shows that the ground scalar meson nonet is composed of the
$a_0(830)$, $f_0(530)$, $f_0(1070)$ and $K^\ast_0(960)$, and thereby suggests that these states
correspond to the observed scalar states $a_0(980)$, $\sigma$, $f_0(980)$ and
$K^\ast_0(930)$\footnote{It is supposed that it is possible for a wide strange resonance,
$K^\ast_0(930)$ to exist in nature still missed in detectors as the ground scalar state whereas the
resonance $K^\ast_0(1430)$ is its radial excitation\cite{9904226}.}, respectively.

 Oller\cite{oller} has already suggested that the $a_0(980)$, $\kappa$, $f_0(980)$ and
$\sigma$ resonances constitute the lightest scalar nonet in three different and complementary ways:
a) by establishing the continuous movement of the poles from the physical to a SU(3) limit, b) by
performing an analysis of the couplings of the scalar mesons to pairs of pseudoscalars and c) by
analysing the couplings of the scalars with meson-meson SU(3) scattering eigenstates. The results
given by Oller\cite{oller} show
\begin{eqnarray}
\left(\begin{array}{c}
f_0(980)\\
\sigma
\end{array}\right)=\left(\begin{array}{ll}
0.28&-0.96\\
0.96&~~0.28
\end{array}\right)\left(\begin{array}{ll}
N\\
S
\end{array}\right).
\label{content1}
\end{eqnarray}
 Clearly, the agreement between (\ref{content}) and
(\ref{content1}) is good.

It is worth mention that our suggested $q\bar{q}$ assignment for the ground scalar nonet is also
favored by the results suggested by U(3)$\times$U(3) $\sigma$ model\cite{nap} and  SU(3) $\sigma$
model\cite{su3}.

Finally, we remark also that the masses of the $f_0(1099)$ and $f_0(530)$ predicted in the present
work are below a typical range of $1730\pm 50\pm 80$ MeV suggested by Lattice QCD calculation for
the ground scalar glueball\cite{lattice}. The masses of the two isoscalar scalar mesons may get
shifted from the predicted values due to the possible mixture with the ground scalar glueball.

\section*{V. Concluding remarks}
\indent\vspace*{-1cm}

In the presence of the $a_0(980)$ and $D^\ast_{sJ}(2317)$ belonging to the $1~^3P_0$ $q\bar{q}$
multiplet, we estimate the mass of the $1~^3P_0$ kaon meson in the framework of the quasi-linear
Regge trajectory. Then in the framework of the
 meson-meson mixing, we suggest the $a_0(980)$, $K^\ast_0(1052)$, $f_0(1099)$ and
 $f_0(530)$ constitute the ground scalar meson nonet. We find that the $f_0(1099)$
 is  mostly strange while the $f_0(530)$ is mainly non-strange. We suppose that
 the $K^\ast_0(1052)$, $f_0(1099)$ and
 $f_0(530)$ would likely correspond to the observed scalar states $\kappa(900)$,
 $f_0(980)$ and $f_0(600)/\sigma$, respectively.
 Our suggested $q\bar{q}$ assignment for the $1~^3P_0$ meson
  nonet is consistent with the assignments established by \cite{nap,su3,9904226,oller} in
  different approaches. The fact that the agreement between the present findings and those
given by other different approaches is satisfactory implies
 that the argument that the $a_0(980)$ and $D^\ast_{sJ}(2317)$ are
 ordinary $1~^3P_0$ $q\bar{q}$ states may be reasonable.

\noindent {\bf Acknowledgments:} This work is supported in part by National Natural Science
Foundation of China under Contract No. 10205012, Henan Provincial Science Foundation for
Outstanding Young Scholar under Contract No. 0412000300, Henan Provincial Natural Science
Foundation under Contract No. 0311010800, and Foundation of the Education Department of Henan
Province under Contract No. 2003140025.

\end{document}